\documentstyle[preprint,aps,tighten]{revtex}
\begin{document}
\draft
\title{
Triton calculations with $\pi$ and $\rho$ exchange three-nucleon forces
}
\author{A. Stadler}
\address{
Institut f\"ur Theoretische Physik, Universit\"at Hannover,
D-30167 Hannover 1, Germany \\
and\\
Department of Physics,
College of William and Mary, Williamsburg, Virginia 23185,
U.S.A.\thanks{Present address.}
}
\author{J. Adam Jr.}
\address{
Institute of Nuclear Physics, CZ-25068, \v{R}e\v{z}, Czech Republic
}
\author{H. Henning, and P.U. Sauer}
\address{
Institut f\"ur Theoretische Physik, Universit\"at Hannover,
D-30167 Hannover 1, Germany}
\date{\today}
\maketitle

\begin{abstract}
The Faddeev equations are solved in momentum space for the trinucleon
bound state with the new Tucson-Melbourne $\pi$
and $\rho$ exchange three-nucleon potentials.
The three-nucleon potentials are combined with a variety of
realistic two-nucleon
potentials. The dependence of the triton binding energy
on the $\pi NN$ cut-off parameter in the three-nucleon potentials is studied
and
found to be reduced compared to the case with pure $\pi$
exchange. The $\rho$ exchange parts of the three-nucleon
potential yield an overall
 repulsive effect. When the recommended
parameters are
employed, the calculated triton binding energy turns out to be
very close to its experimental value. Expectation values of
various components of the three-nucleon potential are given to
illustrate their significance for binding.
\end{abstract}

\narrowtext
\section{INTRODUCTION}

None of the dynamical models for hadronic interactions that have
been constructed in the past to microscopically describe nuclear properties
is fundamental. All models are effective ones;
only the most important hadronic degrees of freedom are taken
into account explicitly. Effective theories in general lead to
rather complicated forces, including irreducible many-body
forces. The complexity and interpretation of these potentials
is closely related to the chosen hadronic degrees of freedom.
Most often the microscopic theory of nuclear phenomena is formulated
in a Hilbert space of nucleons only employing
nonrelativistic quantum mechanics.  The potentials are taken to
be of two-body nature, the mediating meson fields having been
frozen out.  Many-meson exchanges and relativistic effects are
absorbed into the phenomenological short-range part of the potential
that is also introduced to regularize it at the origin.
Many-body forces are assumed to be much less important
than the dominant two-nucleon force. This is a reasonable assumption,
since on average nucleons move relatively slowly
inside nuclei,
three nucleons rarely interact simultaneously due to the short-range
repulsion between two of them, and
the rate of nucleonic excitation into nonnucleonic states is low. The
parameters of these semi-phenomenological models are calibrated
to reproduce the bound-state and scattering observables of the
two-nucleon systems. Therefore,
the three-nucleon system offers the first testing ground for the
two-nucleon potentials. Can a Hamiltonian with only two-nucleon
potentials
successfully describe the trinucleon properties or not? The
qualitative and quantitative effects
of three-nucleon forces must be understood.

The derivation of realistic two-nucleon potentials has been
steadily improved in the past. See the very good review  of
Ref.~\cite{Mac89} on this subject.
Furthermore, the techniques to solve the
three-nucleon bound-state problem have reached maturity,
the vast progress became possible by the rapid advances of
computational power. The trinucleon calculations made it clear
that no realistic  two-nucleon potential is able to reproduce the known
hadronic and e.m.\
properties of $^3{\rm H}$ and $^3{\rm He}$ with satisfactory accuracy.
E.g., the
triton binding energy obtained from such two-nucleon Hamiltonians
turns out to be
about 1 MeV below its experimental value. Calculations  of
$^4{\rm He}$ show a similar deficiency in binding
energy \cite{Car88,Wir91,Kam93}.
Thus, three-nucleon forces have to provide additional
net attraction in order to close the gap between theoretical prediction for
and experimental value of nuclear binding energies.

Two strategies for introducing three-nucleon forces have been investigated:
In the first one a three-nucleon potential is
designed to be added to a conventional
Hamiltonian with two-nucleon potentials
\cite{Coo79,Coo81,Rob86,Ued84,Dei91}. In the second one,
the role of nucleonic excitation in generating
three-nucleon forces is recognized: E.g.,  in Refs.\ \cite{Haj79,Haj83}
three-nucleon forces are
obtained by the explicit inclusion of a $\Delta$ isobar in an
extended Hilbert space. It came as a surprise that
both strategies yield rather different trinucleon binding energies, even when
the
underlying physical processes are thought to be comparable. An
attempt to combine the two approaches has been made in Ref.~\cite{Sta92}
where also their respective advantages and disadvantages are discussed. The
present paper reports on triton calculations with a new three-nucleon force
\cite{Coo93} that belongs to the first strategy.

Like all modern realistic two-nucleon potentials, the models for a
three-nu\-cle\-on force
are based on
meson exchange.  Since nucleons are kept apart by short-range
two-nucleon repulsion,
the two-pion ($\pi\pi$) exchange part of
the three-nucleon force is expected to be dominant. Several  $\pi\pi$
exchange three-nucleon potentials have been worked out.
When used in triton calculations
they provide ample additional binding energy. However, in contrast to
expectation, its effect
is strongly dependent on the  cut-off parameter $\Lambda_{\pi NN}$
entering the regularizing form factor at the $\pi$NN vertices, i.e.,
it is dependent on non-pion physics.
These purely phenomenological form factors are usually taken either in
monopole or in square root parametrization. Since the
same vertex also enters all models of the two-nucleon and the pion-nucleon
potentials, both nucleon-nucleon
and pion-nucleon scattering have been used to extract the value of the
cut-off parameter $\Lambda_{\pi NN}$. Unfortunately, the answer is not
unique. The two-nucleon
 potentials seem to
demand ``hard'' form factors with a $\Lambda_{\pi NN}$  of about 1.2 GeV,
whereas the analysis of $\pi$N scattering favors ``soft'' ones with
$\Lambda_{\pi NN}$ typically around 800 MeV (for a discussion see
\cite{Ued92}).
When such a soft
form factor is employed in a $\pi\pi$ exchange three-nucleon potential,
the triton
turns out to be already overbound. Further increase of $\Lambda_{\pi NN}$
leads to unphysically large binding energies well  before
$\Lambda_{\pi NN}$  reaches the characteristic values for hard form
factors.

It is well known from two-nucleon potentials that the exchange of a
$\rho$ meson cancels to some extent the medium range part of the one-pion
exchange tensor force, providing an important medium range repulsion to
the two-nucleon potential. This cancellation also reduces the
cut-off dependence of the two-nucleon potential. Thus,
  the inclusion of $\rho$  exchange into three-nucleon
force models should lead to less cut-off dependent results and simultaneously
to a smaller overall contribution of the three-nucleon force to the triton
binding energy.  In the context of strategy 2 for the three-nucleon force
this hope was confirmed already a long time ago \cite{Haj83b}.

The $\pi\rho$ and $\rho\rho$ contributions to three-nucleon potentials
 were derived already in \cite{Ell85,Mar80,Rob84,Dei91}. Due to
the enormous technical difficulties it is only now that they are
being included in exact triton calculations.
First results with the Brazilian three-nucleon potential
 were presented in Ref.~\cite{Sas92}.
They display the desired repulsive effect of
the $\rho$  exchange on the triton binding, but do not  study
 its cut-off dependence, since the authors were
only interested in determining parameter combinations that
fit the experimental three-nucleon binding energy. Recently,
Coon and Pe\~na \cite{Coo93} reexamined and improved the  $\pi\rho$
and $\rho\rho$ exchange three-nucleon potential \cite{Ell85}
that was developed by the Tucson-Melbourne
group as an extension of their  $\pi\pi$ exchange three-nucleon potential
\cite{Coo79}.

In this paper we report our results of Faddeev calculations in
momentum space for the triton binding energy, where the
 family of $\pi$ and $\rho$ exchange Tucson-Melbourne three-nucleon potentials
is employed. In Section II the
explicit expressions of these potentials are given.  In Section
III the numerical results are displayed and discussed, followed
by a summary in Section IV.

\section{THE TUCSON-MELBOURNE THREE-NUCLEON POTENTIALS}

The full Tucson-Melbourne $\pi\rho$ and $\rho\rho$ exchange
potentials are given in Ref.~\cite{Coo93}. We do not
comment on the physical origin of the various contributions
but just display the final expressions in momentum space.
For completeness we also include the
$\pi\pi$ exchange part. We take over the notation of Ref.\ \cite{Ell85}.
For the definition of the various occurring
momenta see Figs.~\ref{Fig1}-\ref{Fig3}, where the basic diagrams of the
three-nucleon potentials are shown.

The three-nucleon potentials in momentum space read
\widetext
\begin{equation}
\langle {\bf k}_1' {\bf k}_2' {\bf k}_3' | W |
        {\bf k}_1 {\bf k}_2 {\bf k}_3 \rangle  =
\sum_{\alpha\beta=\pi,\rho} \sum_{i=1}^3
\langle {\bf k}_1' {\bf k}_2' {\bf k}_3' | W_i^{\alpha\beta} |
        {\bf k}_1 {\bf k}_2 {\bf k}_3 \rangle   ,
\end{equation}


\begin{eqnarray}
\lefteqn{
\langle {\bf k}_1' {\bf k}_2' {\bf k}_3' | W_1^{\pi\pi} |
        {\bf k}_1 {\bf k}_2 {\bf k}_3 \rangle  =
 \frac{1}{(2\pi)^6}
\frac{ \delta ( {\bf k}_1'+ {\bf k}_2'+ {\bf k}_3'-
        {\bf k}_1- {\bf k}_2- {\bf k}_3 ) }
     { ({\bf q}^2 + \mu^2) ({\bf q'}^2 + \mu^2) }
R^{\pi\pi}({\bf q}^2,{\bf q'}^2)
}
\nonumber \\
& & \times
(\bbox{\sigma}_2 \cdot {\bf q})
(\bbox{\sigma}_3 \cdot {\bf q'})
\left\{ ( \bbox{\tau}_2 \cdot \bbox{\tau}_3 )
     \left[ a + b {\bf q} \cdot {\bf q'}
           +c({\bf q}^2 + {\bf q'}^2) \right]
+(i \bbox{\tau}_1 \cdot \bbox{\tau}_2 \times \bbox{\tau}_3)
d
(i \bbox{\sigma}_1 \cdot {\bf q} \times {\bf q'}) \right\}
\, , \label{Epipi} \nonumber\\
\end{eqnarray}


\begin{eqnarray}
\lefteqn{
\langle {\bf k}_1' {\bf k}_2' {\bf k}_3' | W_1^{\pi\rho} |
        {\bf k}_1 {\bf k}_2 {\bf k}_3 \rangle  =
- \frac{1}{(2\pi)^6}
\frac{ \delta ( {\bf k}_1'+ {\bf k}_2'+ {\bf k}_3'-
        {\bf k}_1- {\bf k}_2- {\bf k}_3 ) }
     { ({\bf q}^2 + m_{\rho}^2) ({\bf q'}^2 + \mu^2) }
(\bbox{\sigma}_3 \cdot {\bf q'})
}
\vspace{1cm} \nonumber \\
  & \times &
\left\{
- (i \bbox{\tau}_1 \cdot \bbox{\tau}_2 \times \bbox{\tau}_3)
R^{\pi\rho}_{KR}({\bf q}^2,{\bf q'}^2)
(i \bbox{\sigma}_1 \cdot \bbox{\sigma}_2 \times {\bf q})
\right.
\nonumber \\
&  & +
( \bbox{\tau}_2 \cdot \bbox{\tau}_3 )
R^{\pi\rho}_{\Delta^+}({\bf q}^2,{\bf q'}^2)
({\bf q} \times {\bf q'}) \cdot
({\bf q} \times \bbox{\sigma}_2)
\nonumber \\
& & + \left.
  (i \bbox{\tau}_1 \cdot \bbox{\tau}_2 \times \bbox{\tau}_3)
R^{\pi\rho}_{\Delta^-}({\bf q}^2,{\bf q'}^2)
\left[
(i \bbox{\sigma}_1 \cdot \bbox{\sigma}_2 \times {\bf q'}) {\bf q}^2
- (i \bbox{\sigma}_1 \cdot {\bf q} \times {\bf q'})
(\bbox{\sigma}_2 \cdot {\bf q})
\right]
\right\}
\nonumber \\
& & \qquad\qquad + (2 \leftrightarrow 3 \, ,
\qquad {\bf q} \leftrightarrow -{\bf q'})
\, , \label{Epirho}
\end{eqnarray}
and


\begin{eqnarray}
\lefteqn{
\langle {\bf k}_1' {\bf k}_2' {\bf k}_3' | W_1^{\rho\rho} |
        {\bf k}_1 {\bf k}_2 {\bf k}_3 \rangle =
- \frac{1}{(2\pi)^6}
\frac{ \delta ( {\bf k}_1'+ {\bf k}_2'+ {\bf k}_3'-
        {\bf k}_1- {\bf k}_2- {\bf k}_3 ) }
     { ({\bf q}^2 + m_{\rho}^2) ({\bf q'}^2 + m_{\rho}^2) }
}
\nonumber \\
&  \times &
\left\{
(i \bbox{\tau}_1 \cdot \bbox{\tau}_2 \times \bbox{\tau}_3)
R^{\rho\rho}_{Beg}({\bf q}^2,{\bf q'}^2)
i \bbox{\sigma}_1 \cdot
(\bbox{\sigma}_2 \times {\bf q} ) \times
(\bbox{\sigma}_3 \times {\bf q'} )
\right.
\nonumber \\
& & +
( \bbox{\tau}_2 \cdot \bbox{\tau}_3 )
R^{\rho\rho}_{\Delta^+}({\bf q}^2,{\bf q'}^2)
\left[
( \bbox{\sigma}_2 \times {\bf q} ) \times {\bf q}
\cdot
( \bbox{\sigma}_3 \times {\bf q'} ) \times {\bf q'}
\right]
\nonumber \\
& & +
\left.
(i \bbox{\tau}_1 \cdot \bbox{\tau}_2 \times \bbox{\tau}_3)
R^{\rho\rho}_{\Delta^-}({\bf q}^2,{\bf q'}^2)
i \bbox{\sigma}_1 \cdot
\left[
\left(
( \bbox{\sigma}_2 \times {\bf q} ) \times {\bf q})
\times
(( \bbox{\sigma}_3 \times {\bf q'} ) \times {\bf q'}
\right)
\right]
\right\}  \, . \label{Erhorho}
\end{eqnarray}
\narrowtext
The following shorthands have been used in (\ref{Epipi}), (\ref{Epirho}), and
(\ref{Erhorho}):
\widetext
\begin{eqnarray}
R^{\pi\pi}({\bf q}^2,{\bf q'}^2)
& = &
\frac{g^2}{4m^2}
F^2_{\pi NN}({\bf q}^2)
F^2_{\pi NN}({\bf q'}^2) \, ,
\\
R^{\pi\rho}_{KR}({\bf q}^2,{\bf q'}^2)
& = &
\frac{g_\rho^2}{16m^3}
\left[ F_{{\rho NN}_D} ( {\bf q}^2 )  + \kappa_{\rho}
       F_{{\rho NN}_P} ( {\bf q}^2 ) \right]
F_{{\rho NN}_D} ( {\bf q}^2 )
g^2
F^2_{\pi NN} ( {\bf q'}^2 ) \, , \label{Ea-}
\\
R^{\pi\rho}_{\Delta^+}({\bf q}^2,{\bf q'}^2)
& = &
\frac{g_{\rho}}{48m^5}
\left[ F_{{\rho NN}_D} ( {\bf q}^2 )  + \kappa_{\rho}
       F_{{\rho NN}_P} ( {\bf q}^2 ) \right]
\nonumber \\
& & \times
G_{M\rho}^* F_{\rho N \Delta} ( {\bf q}^2 )
\frac{m}{M}
\frac{5M-m}{M-m}
\left[ m g^* F_{\pi N \Delta} ( {\bf q'}^2 ) \right]
F_{\pi NN} ( {\bf q'}^2 ) \, , \label{Ea+b+}
\\
R^{\pi\rho}_{\Delta^-}({\bf q}^2,{\bf q'}^2)
& = &
\frac{1}{4} R^{\pi\rho}_{\Delta^+}({\bf q}^2,{\bf q'}^2) \, , \label{Eprd-}
\\
R^{\rho\rho}_{Beg}({\bf q}^2,{\bf q'}^2)
& = &
\frac{g_\rho^4}{64m^3}
\left[ F_{{\rho NN}_D} ( {\bf q}^2 )  + \kappa_{\rho}
       F_{{\rho NN}_P} ( {\bf q}^2 ) \right]
\nonumber \\
& & \times
\left[ F_{{\rho NN}_D} ( {\bf q'}^2 )  + \kappa_{\rho}
       F_{{\rho NN}_P} ( {\bf q'}^2 ) \right]
\left[ 1+ {\tilde \kappa}_\rho (0) \right]
 F^2_{{\rho NN}_D} (0) \, , \label{Er-}
\\
R^{\rho\rho}_{\Delta^+}({\bf q}^2,{\bf q'}^2)
& = &
\frac{g_\rho^2 \lambda_M^2}{18(M-m)}
\left[ F_{{\rho NN}_D} ( {\bf q}^2 )  + \kappa_{\rho}
       F_{{\rho NN}_P} ( {\bf q}^2 ) \right]
\nonumber \\
& & \times
F_{\rho N \Delta} ( {\bf q}^2 )
\left[ F_{{\rho NN}_D} ( {\bf q'}^2 )  + \kappa_{\rho}
       F_{{\rho NN}_P} ( {\bf q'}^2 ) \right]
F_{\rho N \Delta} ( {\bf q'}^2 ) \, , \label{Er+}
\\
R^{\rho\rho}_{\Delta^-}({\bf q}^2,{\bf q'}^2)
& = &
\frac{1}{4} R^{\rho\rho}_{\Delta^+}({\bf q}^2,{\bf q'}^2) \, , \label{Errd-}
\end{eqnarray}
\narrowtext
with ${\tilde \kappa}_\rho (0) = 3.7 $ and
$\lambda_M = - \frac{3}{2m(M+m)} G^*_{M\rho} $,
where $\mu$, $m_\rho$, $m$, and $M$ denote the pion, the rho,
the nucleon, and the $\Delta$ mass respectively.

Note that the potentials (\ref{Epipi}) -- (\ref{Erhorho})
represent only the part of the full force in which particle 1
interacts both with particles 2 and 3 (see Figs.~\ref{Fig1}-\ref{Fig3}).
The other two contributions with particles 2 and 3 in the middle
of the diagrams are obtained by permutations. The potentials are
given as operators in the three-nucleon spin and isospin space.
The employed three-nucleon basis states are normalized as
\begin{equation}
\langle {\bf k}_1' {\bf k}_2' {\bf k}_3' |
        {\bf k}_1 {\bf k}_2 {\bf k}_3 \rangle =
\delta ( {\bf k}_1' - {\bf k}_1)
\delta ( {\bf k}_2' - {\bf k}_2)
\delta ( {\bf k}_3' - {\bf k}_3) \,  .
\end{equation}

The general structure of the regularization form factors $F_i$
at the meson-baryon-baryon vertices is taken to be of monopole
form, i.e.,
\begin{equation}
F_i({\bf q}^2) = \frac{\Lambda_i^2 -m_b^2}
                       {\Lambda_i^2 + {\bf q}^2 } \, , \label{Eff}
\end{equation}
with $i=\{\pi NN,\pi N \Delta , \rho NN_D , \rho NN_P, \rho N \Delta \} $
and  $m_b$ the mass of the boson at the vertex (i.e., $m_b$ is either
$\mu$ or $m_\rho$; an exception is the case $i=\rho N \Delta$, where
$m_b = 0$), $F_{{\rho NN}_D}$ being the Dirac and $F_{{\rho NN}_P}$
the Pauli form factor.
The numerical values of the employed masses and
potential parameters are
given in Table \ref{Tparameters}.
Note that the values of our parameters $a$, $b$, and $c$, are taken from
Refs.\ \cite{Coo79,Coo81} and differ slightly from those of Ref.\ \cite{Coo93}.
The latter have been extracted from an updated set of experimental $\pi N$
data and were published only after most of our numerical calculations have
been completed
(a short account of our results appeared in Ref.\ \cite{Sta93}, where
unfortunately the $\rho\rho$ part of the three-nucleon force was treated
incorrectly).

Furthermore, the original \cite{Ell85} factor of ${1 \over 4}$
in Eq.\ (\ref{Eprd-}) has been
changed in Ref.\ \cite{Coo93} to ${1 \over 2} \frac{M+m}{5M-m}$ which
numerically is approximately ${1 \over 4.8}$ and also represents only a
small variation. Both changes do not appear significant enough to
justify repeating our very elaborate computations carried out prior to
Ref.\ \cite{Coo93}. However, we have
checked numerically that their net effect in the triton binding energy
is small. More details can be found in Section III.

\section{Results}

The Faddeev equations for the three-nucleon bound state with the
inclusion of an irreducible three-nucleon force can be written as
\cite{Sta91}
\begin{eqnarray}
| \psi_i \rangle & = & G_0(E)
\left\{  T_i(E) P  \right.
\nonumber \\
&  & +
\left. \left[ 1+T_i(E) G_0(E) \right]
W_i (1+P) \right\} | \psi_i \rangle  \, , \label{Efaddeev}
\end{eqnarray}
where $| \psi_i \rangle $ is the Faddeev amplitude, $G_0(E)$
the free three-nucleon propagator, $T_i(E)$ the two-nucleon
transition matrix embedded in the three-nucleon Hilbert space,
$W_i$ the potential operator of the irreducible three-nucleon
force, and $P$ the sum of the two operators for cyclic and
anticyclic permutations of the three particles. The subscript $i$
denotes in $| \psi_i \rangle $ and $T_i(E)$ the spectating
particle, and in $W_i$ the particle that interacts simultaneously
with the other two. $E$ is the  the trinucleon binding energy.
Equation (\ref{Efaddeev}) is employed in momentum space and
expanded into partial waves. The explicit form of the resulting
set of coupled integral equations in two continuous variables as well as
details of the solution method  have been presented
in Ref.~\cite{Sta91}.
All results of this section were obtained in a basis of 18
partial waves. They correspond to all possible combinations of
spin-, isospin- and orbital angular-momentum quantum numbers
that can be coupled to the quantum numbers of the triton, when
the two-nucleon pair interaction is restricted to total
angular momenta up to 2.

The main difficulty of this work as an extension of Ref.~\cite{Sta91}
was the partial wave decomposition of the $\pi \rho$ and $\rho \rho$
exchange
three-nucleon potentials, since their structure is considerably
more complex than that of the $\pi \pi$ exchange potential.
This complexity calls for a systematic procedure to decompose
general three-nucleon operators into partial waves. Such a
procedure was developed by Adam and Henning \cite{Ada92,Hen92}
and is applied in this work for the particular case of the
Tucson-Melbourne three-nucleon potentials.
Given the matrix elements of the three-nucleon potentials in
partial wave decomposed form, their numerical evaluation is
still a formidable task. In order to illustrate the
enormous computing resources necessary for performing such
calculations we note that the computation of the $\pi \pi$ exchange
potential matrix elements in 18 channels on a Siemens/Fujitsu S400
Supercomputer, as it was done for Ref.~\cite{Sta91}, took about
90 minutes of CPU time. Therefore it was absolutely inevitable
to develop a new  method that exploits both the advantages of the
general technique for multipole decomposition and the possibilities
of efficient vectorization as much as possible.
 With that new method, the $\pi \pi$ exchange matrix elements
are calculated within 3.5 minutes, the $\pi \rho$
exchange matrix elements within
10 minutes, and the $\rho \rho$ exchange matrix elements within
15 minutes of CPU time.
The accuracy  of the new code was tested by comparing $\pi \pi$ exchange
potential matrix elements with results from the old  technique
and was found to be improved, too. Triton binding
 energies including the effect of  $\pi \pi$ exchange three-nucleon
potentials calculated with the old and the new codes agree.

Now we turn to the presentation of the triton binding energies
obtained with the Tucson-Melbourne three-nucleon potentials.

In Table~\ref{Tfull} the results for the Reid soft core \cite{Rei68},
Paris \cite{Lac80}, Nijmegen \cite{Nag78}, and Bonn OBEPQ \cite{Mac87}
 as underlying two-nucleon  potentials are given.
We consider these potentials as representative for various
approaches to modeling the two-nucleon  force: they are
purely phenomenological, based on  dispersion theory, non-relativistic
and derived from one-boson exchange, and  derived from one-boson
exchange with minimal relativity, respectively.

The binding energies obtained with the complete $\pi$ and
$\rho$ exchange  Tucson-Melbourne
three-nucleon potentials (which we hereafter shorthand as
``full results'') scatter closely around the experimental triton
binding energy of -8.482 MeV. The only exception occurs in the case of
 the Bonn OBEPQ potential, which comes close to that value
already without any three-nucleon force. The full result for the
Paris potential happens to deviate from the experimental value by only
12 keV.

As expected, the $\pi \rho$ three-nucleon potential decreases the
trinucleon binding energy, compensating to a large extent the
overbinding due to the $\pi \pi$ potential alone. We also observe
that the $\rho\rho$ potential has only a small effect.
For the $\Delta$
mediated three-nucleon force of the Hannover model, it was found
already many years ago \cite{Haj83b} that the $\pi\rho$ exchange
weakens the attraction due to the $\pi\pi$ exchange and that the
contribution of the  $\rho\rho$ exchange is rather small.
These findings are also in good qualitative agreement
with the results obtained by Sasakawa {\em et al.\ } \cite{Sas92}
for the simpler Brazilian three-nucleon potentials.
However, we want to point out that the calculations of Ref.\ \cite{Sas92}
were performed with a different two-nucleon potential, that they include
only the $\Delta$ part of the $\rho\rho$ potential, and that the implemented
functional form of the $\rho$ form factors as well as the values of the $\rho$
cut-off parameters are very different from ours.
That the effect of the $\rho\rho$ potential in our calculations can be
attractive or repulsive, depending on the employed two-nucleon
potential, while Ref.\ \cite{Sas92} reports only repulsion,
is therefore not inconsistent.

However, one should not take the impressive agreement of theoretical binding
energies with experiment
too seriously, even though no attempt was made to
adjust any parameters of the  three-nucleon potential in order to reproduce
the experimental trinucleon binding energy.
One should keep in mind that the exact values of the cut-off
parameters $\Lambda_i$ are not well determined, although some arguments have
been given \cite{Coo93} favoring the set of parameters  listed in
Table~\ref{Tparameters}.
In particular, the recommended value for the $\pi NN$ vertex,
$\Lambda_{\pi NN} = 5.8 \mu \approx 810$ MeV, is based on
an analysis of the Goldberger-Treiman relation \cite{Coo90}. The
uncertainties associated with the extrapolation of the physical
constants in the Goldberger-Treiman relation to their chiral-limit
values are estimated in Ref.\ \cite{Coo93} to yield a ``theoretical
error bar'' of about $\pm 200$ MeV for $\Lambda_{\pi NN}$,
an uncertainty which creates substantial variation in the calculated
triton binding energy.

Since it is known already from previous studies with the
$\pi\pi$ exchange three-nucleon force \cite{Sas86,Che86,Sta91} that
the results are strongly dependent on the cut-off $\Lambda_{\pi NN}$
it is certainly necessary
to reexamine that dependence for the full three-nucleon force.
Each full calculation still requiring  enormous computational
capacities, we had to restrict ourselves to a small number of combinations
of the cut-off parameters. Nevertheless, the studied variation  is sufficient
to draw some qualitative conclusions.

The variation of $\Lambda_{\pi NN}$ is performed using the same
values as before in the study of the $\pi \pi$ exchange three-nucleon
force \cite{Che86,Sta91}. For the  cut-off parameters
connected with the $\rho$ meson we consider just  two cases,
i.e.,  on one hand soft form factors, represented by the Tucson-Melbourne
 choice $\Lambda_{{\rho NN}_D} = 12.0 \mu$ and $\Lambda_{{\rho NN}_P} =
  7.4 \mu$, and on the other hand  hard form factors
 as suggested by Ref.~\cite{Mar80}, with
$\Lambda_{{\rho NN}_D} = \Lambda_{{\rho NN}_P} =
\Lambda_{\rho N\Delta} = 13.4 \mu$. These two choices of $\rho$
cut-off parameters have also been considered in Ref.~\cite{Coo93}.

The results with the Paris potential as two-nucleon potential are shown
in Table \ref{Tcutoff} and in Figure 4.
The strong dependence on $\Lambda_{\pi NN}$ is considerably reduced  once
the $\pi\rho$ part is added to the three-nucleon force,
 although the remaining dependence is
still  sizeable. Changing the $\rho $ cut-off
parameters from ``soft'' to ``hard'' values also alters
the triton binding energy: The effect of the $\rho\rho$ part
is significantly enhanced for hard form factors. It is repulsive and greater
than
200 keV (for $\Lambda_{\pi NN} = 5.8 \mu$), whereas it is less than 10 keV
and attractive for soft form factors.

However, there are indications that
the use of hard $\rho$ cut-off parameters
in the Tucson-Melbourne potential does not lead to physically meaningful
results.
Although our primary reference two-nucleon potential is the Paris potential, we
repeated a calculation with hard  $\rho $ cut-off parameters also for the Reid
potential.
We found that the $\rho\rho$ potential becomes so strong that no converged
trinucleon
binding energy could be obtained (for $\Lambda_{\pi NN} = 5.8 \mu$).
Apparently,
there
are strong and model-dependent cancellations between different components of
the
$\rho\rho$ potential which happen to reduce to a comparably small net result in
the
case of the Paris potential, while for other potentials they can diverge.
A similar problem occurs already when only the $\pi\pi$ potential is
considered:
While stable trinucleon binding energies for the Reid and Paris potentials can
be obtained
when $\Lambda_{\pi NN} = 7.1 \mu$, they diverge for Nijmegen or Bonn
potentials.

We see two possible interpretations of these divergencies for hard cut-off
parameters.
Either the Tucson-Melbourne parametrization of the three-nucleon force is valid
and
there are physical reasons to favor soft form factors. Or the Tucson-Melbourne
potentials
are {\em mathematically} not well defined for hard form factors, i.e., in the
course of
solving the three-nucleon equations the Tucson-Melbourne potentials are
evaluated outside
their range of convergence associated with various Taylor expansions that are
employed in their
derivation. We believe it would be important to further investigate these
questions. However,
it is not the purpose of this paper, in which we rather want to focus on the
original parametrization
of the Tucson-Melbourne potentials.

The use of the updated parameters of the Tucson-Melbourne potential
 instead of the ones of
Table \ref{Tparameters} changes the triton binding energy by very little.
The check was carried out by a complete recalculation for the entry
-8.494 MeV under $\pi\pi+\pi\rho+\rho\rho$ soft with $\Lambda_{\pi NN} = 5.8
\mu$
for the Paris potential in Table \ref{Tcutoff}.
The updated parameters $a$, $b$,
and $c$, in the $\pi \pi$ part of the three-nucleon potential change the
triton binding energy from -8.494 MeV
to -8.465 MeV. The subsequent modification of the factor ${1 \over 4}$ in
the isospin-odd
$\Delta$ part of the $\pi\rho$ potential in (\ref{Eprd-})
to the
new prescription, discussed at the end of Sect.\ II, changes the value
of  -8.465 MeV to -8.492 MeV.
Clearly, the found changes are small, in particular if compared to the
sensitivity
of the results to variations of the cut-off parameters.

Table \ref{Texpect} lists expectation values of the various components
of the Tucson-Melbourne three-nucleon potentials. These components
are terms of different spin-, isospin-, and momentum dependence. In
the $\pi\pi$ part of the three-mucleon potential they are labeled
 by their respective coefficients
$a$, $b$, $c$, and $d$. In the case of the $\pi\rho$ and $\rho\rho$
potential they originate from distinct physical processes and are
characterized accordingly as $\Delta$, $K.R.$, and $Beg$. Here $\Delta$
stands for the contributions originating from the $\Delta$ resonance,
whereas $K.R.$ and $Beg$ refer to the terms obtained from the Kroll-Ruderman
and B\'eg low-energy theorems \cite{Ell85}. The superscripts $(\pm)$
indicate their isospin symmetry under particle exchange.

The expectation values are calculated
with the fully correlated wave functions, i.e., they are obtained from the
full Hamiltonian including the full three-nucleon potentials. They represent
therefore true expectation values and not results from first order
perturbation theory. In fact, a comparison of the expectation values
with the corresponding binding energy differences demonstrates again the
non-perturbative character of these calculations.

It can be seen that the unproportionally strong over\-binding for the
OBEPQ potential is mainly caused by the $c$-term of the $\pi\pi$ potential.
There seems to be a correlation between the magnitudes of the expectation
values and the probabilities of
the wave function components. As one moves from left to right in
Table \ref{Texpect}, the triton $D$ state probabilities decrease whereas the
absolute values of the total $\pi\pi$, $\pi\rho$, and $\rho\rho$
contributions increase. The same trend holds also for most of the
individual contributions.

In the $\pi\pi$ part of the three-nucleon potential,
 several physical processes lead to terms of
the same spin-isospin and momentum structure. They are usually not
separated in order to emphasize the model-independent character
of the potential. However, it is interesting to see how the three-nucleon
potential
is composed of physical processes, even if a certain model dependence is
introduced.
In Ref.\ \cite{Coo93} the four coefficients of the $\pi\pi$ potential
are split into different parts according to their physical origin.
There, the given numerical values are based on the updated set of parameters.
Since we are working with the original set of parameters, we have to
recalculate this splitting accordingly. This is the reason why our
values (\ref{Ed}) are not identical to those of \cite{Coo93}.

The coefficients are decomposed as
\begin{eqnarray}
a & = & a_\sigma  \nonumber\\
b & = & b_\Delta +  b_\sigma \nonumber\\
c & = & c_\sigma + c_Z \nonumber\\
d & = & d_\Delta + d_Z + d_{ca}  \, ,
\label{Eabcd}
\end{eqnarray}
with the numerical values
\begin{eqnarray}
a_\sigma & = & 1.13 \mu^{-1}  \nonumber\\
b_\Delta & = & -1.676 \mu^{-3} \, , \; \;
b_\sigma = -0.904 \mu^{-3}  \nonumber\\
c_\sigma & = & 1.15 \mu^{-3} \, , \;\;
c_Z = -0.15 \mu^{-3}  \nonumber\\
d_\Delta & = & -0.36 \mu^{-3}  \, , \;\;
d_Z = -0.15 \mu^{-3} \, , \;\;
d_{ca} = -0.243 \mu^{-3}\,  . \label{Ed}\nonumber\\
\end{eqnarray}
The subscripts $\Delta$, $\sigma$, and $Z$, refer to $\Delta$-isobar,
$\sigma$-term, and Z-graph contributions; $ca$ is a ``current algebra''
term \cite{Coo93}.   It represents a (isovector) vector-current contribution
to the $\pi N$ amplitude which the three-nucleon potential is based on.
In a simple vector-meson dominance model it reduces to a t-channel
$\rho$ exchange part of the $\pi N$ amplitude.

In this context it may be worth mentioning that not only the extraction
of $\Delta$ contributions is model dependent, but also the separation
into Z-graph and $\sigma$-term. In the Tucson-Melbourne $\pi N$ amplitude,
pure pseudoscalar $\pi NN$ coupling is assumed. The separation into
Z-graph and $\sigma$-term would change if pseudovector coupling
were introduced.

The expectation values of Table \ref{Texpect} are reorganized according
to this analysis and displayed in Table \ref{Tsplit}.  The upper half contains
only the results from the $\pi\pi$ potential, the lower half adds
the respective pieces from the $\pi\rho$ and $\rho\rho$ potentials. The
Kroll-Ruderman term is part of what is called Z-graph contributions, because
as in the $\pi\pi$ potential pseudoscalar $\pi NN$ coupling is assumed.
The  B\'eg term is counted as a ``current algebra'' contribution, since it
originates from t-channel 3$\rho$ and $\rho\rho NN$ contact terms.
The $\sigma$-term appears only in the
$\pi\pi$ potential. The corresponding expectation values are therefore
identical
in the upper and lower half of  the table.

The $\Delta$ and $\sigma$-term contributions are both attractive and comparable
in magnitude in the $\pi\pi$ potential. In the $\pi\rho$ potential, however,
the
$\Delta$
parts yield relatively strong repulsion, thus canceling to some extent the
attraction
they produce in the $\pi\pi$ potential. The Z-graph expectation values turn out
to
be small and repulsive, most of which is due to the Kroll-Ruderman term in the
$\pi\rho$ potential.  The current algebra term is similar in magnitude but
opposite
in sign. Again, the results obtained with the OBEPQ potential take on rather
extreme
values. For the Paris potential, the expectation values of the $\Delta$ parts
are quite similar to
the ones obtained with the $\Delta$ mediated three-nucleon force presented in
table 3 of Ref.\ \cite{Haj83b}.

\section{Summary}

We have solved the Faddeev equations  with a
new set of Tucson-Melbourne three-nucleon potentials, based on $\pi$ and
$\rho$ exchange,  together with realistic
two-nucleon potentials for the three-nucleon bound state.
We find that the $\pi\rho$ exchange three-nucleon potential is repulsive and
counteracts the too strong attraction generated by the $\pi\pi$ potential to
such
an extent that the resulting triton binding energies for the Reid soft core,
the
Paris, and
the Nijmegen potentials are close to the experimental
value. This is not true for the Bonn OBEPQ potential which yields overbinding
of more than 1 MeV.
The strong dependence of the binding energy on the $\pi NN$ vertex cutoff
parameter $\Lambda_{\pi NN}$ that was observed for the
Tucson-Melbourne $\pi\pi$ exchange  three-nucleon potential is reduced
once the $\rho$ exchange is added.
The $\rho\rho$ exchange potential has only a small effect on the triton
binding energy as long as the vertex form factors are
not chosen to be  ``hard''.

We have calculated expectation values of the various components of the
three-nucleon potentials. They are presented in two different ways: first
they are grouped according to their spin-isospin and
momentum dependence, and second they are rearranged to exhibit
the relative strength of the underlying physical processes. We find that
the contributions from the $\sigma$-term and from $\Delta$ excitations
are dominant and attractive, compared to which the repulsive effect
of intermediate Z-graphs is small. It is the $\sigma$-term part of the
$\pi\pi$  potential that  appears to be very sensitive to details of the
three-nucleon wave function and that leads to the unusually strong overbinding
in the case of the Bonn OBEPQ potential.

In the derivation of three-nucleon potentials, vertex functions are usually
expanded in powers of particle momenta. The
$b$, $c$, and $d$ parts of the $\pi\pi$  potential
are terms two orders higher in the pion momenta than the $a$ term.
The fact that they yield
considerably larger expectation values in the triton
might indicate a failure of the momentum expansion.  Such a failure could
simultaneously be responsible for the observed strong dependence on the $\pi$
and
$\rho$ cut-off parameters.

That the effect of the $\pi\rho$ three-nucleon potential, although
already much smaller than that of the $\pi\pi$ potential,  is still of
non-negligible size
suggests that other exchange processes involving heavier mesons, such as
$\pi\sigma$ and $\pi\omega$ exchange, should also be investigated.
Since $\sigma$ and $\omega$ exchange are included in most one-boson
exchange two-nucleon potentials, they should also be included in three-nucleon
potentials already for reasons of consistency. These processes have been
found to be important in $\pi$ production on two-nucleon
systems at threshold (see, e.g., Ref.\ \cite{Hor93,Lee93}).

We thank S.\ A.\ Coon and M.\ T.\ Pe\~na for many helpful discussions on
details
of the Tucson-Melbourne three-nucleon potentials.
This work was funded by the Deutsche Forschungsgemeinschaft (DFG) under the
Contract Nos.\ \hbox{Sa 247/7-2}, Sa 247/7-3, Sa 247/9-4 and 436 CSR-111/4/90,
and by the DOE under Grant No.\ DE-FG05-88ER40435. During part of the work,
J.\ A.\ was fellow of the Humboldt Foundation. The calculations were performed
at Regionales Rechenzentrum f\"ur Niedersachsen (RRZN), Hannover, at
Rechenzentrum Kiel, at Continuous Electron Beam Accelerator Facility (CEBAF),
and at National Energy Research Supercomputer Center (NERSC), Livermore.

%
%
\begin{table}
\setdec 000.0000
\caption{Parameters and constants of the Tucson-Mel\-bourne
three-nucleon potential.}
\begin{tabular}{l  c}
 Parameter & Value \\
 \tableline
 $a$   &  \dec 1.13 $\mu^{-1} $\\
 $b$   &  \dec -2.58 $\mu^{-3}$ \\
 $c$   &   \dec 1.00 $\mu^{-3}$ \\
 $d$ & \dec -0.753 $\mu^{-3}$ \\
\tableline
 $m$    &  \dec 938.92 MeV \\
 $M$    &  \dec  1232.  MeV  \\
 $\mu$  &  \dec  139.6 MeV  \\
 $m_\rho$  & \dec 768.3 MeV \\
 \tableline
 $g$    &    $\sqrt{179.7} $  \\
 $g_\rho$  & \dec 5.3  \\
 $\kappa_\rho$ & \dec 6.6  \\
 $g^*$    &  \dec 1.82 $\mu^{-1}$ \\
 $G^*_{M\rho} F_{\rho N \Delta}(0)$ & \dec 14.7 \\
\tableline
$\Lambda_{\pi NN} $     &  \dec  5.8 $\mu$ \\
$\Lambda_{\pi N \Delta}$  & \dec 5.8 $\mu$ \\
$\Lambda_{{\rho NN}_D}$   & \dec  12.0 $\mu$ \\
$\Lambda_{{\rho NN}_P}$    & \dec 7.4 $\mu$ \\
$\Lambda_{\rho N \Delta}$  & \dec 5.8 $\mu$ \\
\end{tabular}
\label{Tparameters}
\end{table}
%
%
\begin{table}
\setdec 0.000
\caption{Triton binding energies in MeV calculated
for different combinations
of two-nucleon and three-nucleon potentials. The column labeled
``no 3NP'' shows the results when no three-nucleon potential is
included. The parameter set of Table \protect\ref{Tparameters} is employed.}
\begin{tabular}{l c c c c }
     &  no 3NP  &  $\pi\pi $ &  $\pi\pi + \pi\rho $ &
$\pi\pi + \pi\rho + \rho\rho $ \\
\tableline
RSC &  \dec -7.229  & \dec -8.904  & \dec -8.438  & \dec -8.451 \\
Paris          & \dec  -7.381  & \dec -9.060  & \dec -8.486  & \dec -8.494 \\
Nijmegen       & \dec  -7.537  & \dec -9.347  & \dec -8.692  & \dec -8.689 \\
OBEPQ     &  \dec -8.315  & \dec -11.056  & \dec -9.639  & \dec -9.596 \\
\end{tabular}
\label{Tfull}
\end{table}
%
%
\begin{table}
\setdec 0.000
\caption{Triton binding energies in MeV calculated for various cut-off
parameters in the employed three-nucleon potentials. For the
$\rho$-baryon-baryon vertices, the set of cut-off parameters
of Table \protect\ref{Tparameters} is labeled as ``soft'', whereas
the choice
$\Lambda_{{\rho NN}_D} = \Lambda_{{\rho NN}_P} =
\Lambda_{\rho N\Delta} = 13.4 \mu$ is referred to as ``hard'', thereby
characterizing the corresponding vertex form factors.
The Paris potential was taken as two-nucleon potential.}
\begin{tabular}{l c c c}
  &  \multicolumn{3}{c}{$\Lambda_{\pi NN}$} \\
 three-nucleon potential &  $4.1 \mu$  &  $5.8 \mu $ & $7.1 \mu $ \\
 \tableline
 $\pi\pi$     & \dec -7.543  & \dec -9.060 & \dec -12.313 \\
 \tableline
 $\pi\pi + \pi\rho$ soft & \dec -7.409 & \dec -8.486 & \dec -10.558 \\
 $\pi\pi + \pi\rho$ hard & \dec -7.484 & \dec -8.468 & \dec -10.583 \\
 \tableline
 $\pi\pi + \pi\rho + \rho\rho$ soft & \dec -7.416 & \dec -8.494 &
 \dec -10.558 \\
 $\pi\pi + \pi\rho + \rho\rho$ hard & \dec -7.468 & \dec -8.256 &
 \dec -9.741 \\
 \end{tabular}
 \label{Tcutoff}
 \end{table}
%
%
\begin{table}
\setdec 00.000
\caption{Expectation values of various components of the
Tucson-Melbourne three-nucleon potentials, evaluated with
the respective wave functions fully correlated by two- and
three-nucleon forces. The parameter
set of Table \protect\ref{Tparameters} is used in the calculations.
In the four columns, the three-nucleon potentials
are combined with different two-nucleon potentials.
The superscript $(+)$ or $(-)$ indicates the even or odd symmetry of
the respective potential component under isospin exchange
of two nucleons.
In addition,
the probabilities of $S$, $S'$, $P$, and $D$ waves in the
triton wave function in percent and the corresponding binding
energy $E_T$ are displayed.
The binding energy row $E_T$ is identical to the last column of
Table \protect\ref{Tfull}.
All energies are in MeV.}
\begin{tabular}{l c c c c}
    &  RSC  &  Paris  &  Nijmegen  &  OBEPQ  \\
\tableline
$\pi\pi$  [MeV]& & & &  \\
 $a^{(+)}$  & \dec 0.009 & \dec -0.005 & \dec -0.018 & \dec -0.089 \\
 $b^{(+)}$  & \dec -1.397 & \dec -1.331 & \dec -1.246 & \dec -0.867 \\
 $c^{(+)}$  & \dec -0.432 & \dec -0.661 & \dec -0.912 & \dec -2.313 \\
 $d^{(-)}$  & \dec -0.355 & \dec -0.322 & \dec -0.335 & \dec -0.754 \\
total $\pi\pi$ & \dec -2.174 & \dec -2.319 & \dec -2.511 & \dec -4.024 \\
\tableline
$\pi\rho$  [MeV]& & & &  \\
 ${\rm K.R.}^{(-)}$  & \dec 0.089 & \dec 0.136 & \dec 0.171 & \dec 0.488 \\
 $\Delta^{(+)}$  & \dec 0.169 & \dec 0.151 & \dec 0.159 & \dec 0.319 \\
 $\Delta^{(-)}$  & \dec 0.127 & \dec 0.165 & \dec 0.171 & \dec 0.164 \\
total $\pi\rho$ & \dec 0.385 & \dec 0.452 & \dec 0.501 & \dec 0.970 \\
\tableline
$\rho\rho$ [MeV] & & & &  \\
 ${\rm Beg}^{(-)}$  & \dec -0.002 & \dec 0.008 & \dec 0.017 & \dec 0.050 \\
 $\Delta^{(+)}$  & \dec -0.001 & \dec -0.010 & \dec -0.016 & \dec -0.051 \\
 $\Delta^{(-)}$  & \dec -0.012 & \dec -0.008 & \dec -0.007 & \dec 0.000 \\
total $\rho\rho$ & \dec -0.014 & \dec -0.011 & \dec -0.006 & \dec -0.001 \\
\tableline
total 3NP [MeV]& \dec -1.803 & \dec -1.878 & \dec -2.016 & \dec -3.055 \\
\tableline
 $E_T$   [MeV]   & \dec -8.451 & \dec -8.494 & \dec -8.689 & \dec -9.596 \\
 $P(S)$     & \dec 88.70 & \dec 89.92 & \dec 90.73 & \dec  92.70 \\
 $P(S')$      & \dec 1.12 & \dec 1.10 & \dec  0.90 & \dec 1.00 \\
 $P(P)$     & \dec 0.16 & \dec 0.13 & \dec 0.13 & \dec  0.10 \\
 $P(D)$      & \dec 10.03 & \dec 8.84 & \dec 8.25 & \dec 6.20 \\
\end{tabular}
\label{Texpect}
\end{table}
%
%
\begin{table}
\setdec 00.00
\caption{Expectation values of the
Tucson-Melbourne three-nucleon potential, split into
$\Delta$, $\sigma$-term, Z-graph, and ``current algebra'' contributions.
The energies  are in MeV.}
\begin{tabular}{l c c c c}
    &  RSC  &  Paris  &  Nijmegen  &  OBEPQ  \\
\tableline
$\pi\pi$  [MeV]& & & &  \\
 $\Delta$  & \dec -1.08 & \dec -1.02 & \dec -0.97 & \dec -0.92 \\
 $\sigma$  & \dec -0.98 & \dec -1.23 & \dec -1.50 & \dec -3.05 \\
 Z  & \dec -0.01 & \dec 0.04 & \dec 0.07 & \dec 0.20 \\
 ca  & \dec -0.12 & \dec -0.10 & \dec -0.11 & \dec -0.24 \\
\tableline
total 3NP  [MeV]& & & &  \\
 $\Delta$  & \dec -0.79 & \dec -0.72 & \dec -0.66 & \dec -0.49 \\
 $\sigma$  & \dec -0.98 & \dec -1.23 & \dec -1.50 & \dec -3.05 \\
 Z  & \dec 0.08 & \dec 0.17 & \dec 0.24 & \dec 0.69 \\
 ca  & \dec -0.12 & \dec -0.10 & \dec -0.09 & \dec -0.19 \\
\end{tabular}
\label{Tsplit}
\end{table}
\begin{figure}
\caption{Feynman diagram for the $\pi \pi$ exchange three-nucleon
potential.}\label{Fig1}
\end{figure}
\begin{figure}
\caption{Feynman diagram for the $\pi \rho$ exchange three-nucleon
potential.}\label{Fig2}
\end{figure}
\begin{figure}
\caption{Feynman diagram for the $\rho \rho$ exchange three-nucleon
potential.}\label{Fig3}
\end{figure}
\begin{figure}
\caption[99]{Dependence of calculated triton binding energies on
\protect$\Lambda_{\pi NN}$
for the Tucson-Melbourne three-nucleon potential in combination with the
Paris potential. The horizontal lines represent the experimental triton
binding energy (dotted) and the calculated value without three-nucleon force
(long dashed).
The triangles are calculated binding energies with the $\pi\pi$ potential, the
full (open) circles are binding energies with the full three-nucleon force
with soft  (hard) $\rho NN$ form factors. The lines through
the symbols are drawn to guide the eye. }
\end{figure}


\begin{references}
\bibitem{Mac89} R. Machleidt, Adv. Nucl. Phys. {\bf 19},
                189 (1989).
\bibitem{Car88} J. Carlson,
                Phys. Rev. C{\bf 38}, 1879 (1988).
\bibitem{Wir91} B. Wiringa,
                Phys. Rev. C{\bf 43}, 1585 (1991).
\bibitem{Kam93} W. Gl\"ockle and H. Kamada, Phys.\ Rev.\ Lett. {\bf 71}, 971
(1993).
\bibitem{Coo79} S. A. Coon, M. D. Scadron, P. C. McNamee, B. R. Barrett,
                D. W. E. Blatt, and B. H. J. McKellar, Nucl. Phys. {\bf A317},
                242 (1979).
\bibitem{Coo81} S. A. Coon and W. Gl\"ockle,
 Phys. Rev. C{\bf 23}, 1790 (1981).
\bibitem{Rob86} M. R. Robilotta and H. T. Coelho, Nucl. Phys. {\bf A460},
                645 (1986).
\bibitem{Ued84} T. Ueda, T. Sawada, T. Sasakawa, and S. Ishikawa,
                Prog. Theor. Phys. {\bf 72}, 860 (1984).
\bibitem{Dei91} S. Deister, M. F. Gari, W. Kr\"umpelmann, and
                M. Mahlke, Few-Body Syst. {\bf 10}, 1 (1991).
\bibitem{Haj79} Ch. Hajduk and P. U. Sauer, Nucl. Phys. {\bf A322}, 329 (1979).
\bibitem{Haj83} Ch. Hajduk, P. U. Sauer and W. Strueve,
                Nucl. Phys. {\bf A405}, 581 (1983).
 \bibitem{Haj83b} Ch. Hajduk, P.U.Sauer, Shin Nan Yang,
                Nucl. Phys. {\bf A405},  605 (1983).
\bibitem{Sta92} A. Stadler and P. U. Sauer,
                Phys. Rev. C{\bf 46}, 64 (1992).
\bibitem{Coo93} S. A. Coon and M. T. Pe\~na, Phys. Rev. C{\bf 48}, 2559 (1993).
\bibitem{Ued92} T. Ueda, Phys. Rev. Lett. {\bf 68}, 142 (1992); G. Jansen, J.
W.
Durso, K. Holinde,
                B. C. Pearce, and J. Speth, Phys. Rev. Lett. {\bf 71}, 1978
(1993); A. W. Thomas and
                K. Holinde, Phys. Rev. Lett. {\bf 63}, 2025 (1989).
\bibitem{Ell85} R. G. Ellis, S. A. Coon, and B. H. J. McKellar,
                Nucl. Phys. {\bf A438}, 631 (1985).
\bibitem{Mar80} M. Martzolff, B. Loiseau, P. Grang\'e, Phys. Lett. {\bf 92B},
                46 (1980).
\bibitem{Rob84} M. R. Robilotta and M. A. Isidro Filho, Nucl. Phys. {\bf A414},
                394 (1984).
\bibitem{Sas92} T. Sasakawa, S. Ishikawa, Y. Wu, and T-Y. Saito,
                Phys. Rev. Lett. {\bf 68}, 3503 (1992).
\bibitem{Sta93}  A.\ Stadler, J.\ Adam Jr., H.\ Henning and P.\ U.\ Sauer,
     in {\em Conference Handbook of XIV European Conference
     on Few-Body Problems in Physics}, Amsterdam, 23-27 August, 1993,
     ed. L.\ P.\ Koch, pp.\ 192-193.
\bibitem{Sta91} A. Stadler, W. Gl\"ockle, and P. U. Sauer,
                Phys. Rev. C{\bf 44}, 2319 (1991).
\bibitem{Ada92} H. Henning, J. Adam~Jr., and P. U. Sauer, in preparation.
\bibitem{Hen92} H. Henning, Ph.D. thesis, in preparation.
\bibitem{Rei68} R. V. Reid, Ann. Phys. (N.Y.) {\bf 50}, 411 (1968).
\bibitem{Lac80} M. Lacombe, B. Loiseau, J. M. Richard, R. Vinh Mau,
                J. C\^ot\'e, P. Pir\`es, and R. de Tourreil,
                Phys. Rev. C{\bf 21}, 861 (1980).
\bibitem{Nag78} M. M. Nagels, T. A. Rijken, and J. J. de Swart,
                Phys. Rev. D{\bf 17}, 768 (1978);
                T. A. Rijken, R. A. M. Klomp, and J. J. de Swart,
                Nijmegen preprint THEF-NYM-91.05.
\bibitem{Mac87} R. Machleidt, K. Holinde, and Ch. Elster, Phys. Rep. {\bf 149},
                1 (1987).
\bibitem{Coo90} S. A. Coon and M. D. Scadron, Phys. Rev. C{\bf 42}, 2256
(1990).
\bibitem{Sas86} T. Sasakawa and S. Ishikawa, Few-Body Syst. {\bf 1},
                3 (1986).
\bibitem{Che86} C. R. Chen, G. L. Payne, J. L. Friar, and B. F. Gibson,
                Phys. Rev. C{\bf 33}, 1740 (1986).
\bibitem{Hor93} C. J. Horowitz, Phys. Rev. C{\bf 48}, 2920 (1993).
\bibitem{Lee93} T.-S. H. Lee and D. O. Riska, Phys. Rev. Lett. {\bf 70}, 2237
(1993).
\end{references}
\end{document}